\documentclass[aps,prb,twocolumn]{revtex4}
\usepackage{makeidx}
\usepackage{amsmath,amssymb,amsfonts,amsthm}
\usepackage{graphicx,bm}

\begin{document}

\title{Rotational instability of the electric polarization and divergence of
the shear elastic compliance}
\date{}
\author{F. Cordero,$^{1}$ H.T. Langhammer,$^{2}$ T. M\"{u}ller,$^{2}$ V.
Buscaglia$^{3}$ and P. Nanni$^{3}$}
\affiliation{$^1$ CNR-ISC, Istituto dei Sistemi Complessi, Area della Ricerca di Roma -
Tor Vergata,\\
Via del Fosso del Cavaliere 100, I-00133 Roma, Italy}
\affiliation{$^{2}$ Institute of Chemistry, Martin Luther University Halle-Wittenberg,
Kurt Mothes Stra\ss e 2, 06120 Halle, Germany}
\affiliation{$^{3}$ CNR-IENI, Institute for Energetics and Interphases, Department of
Genoa, via De Marini 6, I-16149 Genova, Italy}

\begin{abstract}
The rotational instability of the electric polarization $\mathbf{P}$ during
phase transformations between ferroelectric phases is of great practical
interest, since it may be accompanied by extremely large values of the
piezoelectric coefficient, and a divergence of the coupled shear compliance
contributes to such enhancements. In the literature, this had been
explicitly calculated in the framework of the Landau theory and discussed
with specific numerical simulations involving tetragonal, orthorhombic and
rhombohedral ferroelectric phases. When monoclinic phases are involved, such
an approach is practically impossible, and an approximated treatment had
been proposed, based on the observation that in those cases there are shear
strains almost linearly coupled to the transverse component of $\mathbf{P}$,
implying a divergence of the Curie-Weiss type in the associated compliances.
Here the argument is extended to the general case of transitions whose major
effect is a rotation of the polarization, and the limits of its validity are
discussed. As experimental verification, the elastic response of BaTiO$_{3}$
is measured and analyzed, together with those of other ferroelectric
perovskites available in the literature, like KNN.
\end{abstract}

\maketitle


\section{Introduction}

Considerable attention in the field of piezoelectric materials is devoted to
the rotational or transverse instability of the polarization \cite%
{FC00,IOI01,IOI02,KPB06,Dam09,ZLX12,SJR14} and, since the piezoelectric
response is made up of dielectric and elastic contributions \cite%
{NH65,Dam06,Cor15}, the elastic response contains important information,\cite%
{SCN13,CCD14b} that is similar and complementary to the piezoelectric
constants. In addition, the elastic properties can be measured also in
non-ferroelectric phases and are practically insensitive to the presence of
free charge carriers from ionized defects, when measured on unpoled ceramic
samples.\cite{note} On the other hand, the elastic response at transitions
between ferroelectric (FE)\ phases cannot be expressed in a simple and
general manner, and only in few instances explicit expressions of the
elastic anomalies have been provided,\cite{Dev51,II99b,II05b} and discussed
on the basis of numerical examples. In fact, while these expressions can be
written in reasonably transparent forms in terms of the spontaneous
polarization $P_{s}$, they become quite complicated when explicitly written
as a function of temperature. In addition, although the derivation of these
expressions in the framework of the Landau theory of phase transitions is
not conceptually difficult, the tedious algebra does not help in gaining an
intuitive physical picture. Therefore, the numerical curves of the
compliances $s_{ij}\left( T\right) $ exhibit divergences where one
intuitively expects them, for example $s_{44}$ and $s_{55}$ of BaTiO$_{3}$
increase rapidly when the orthorhombic (O) phase is approached from the tetragonal (T)
phase,\cite{Dev51} in accordance with the shear deformation of the cell
transforming from the T to the O structure,\cite{Dev51} and
similar enhancements are found when the anisotropy of the free energy versus
the direction of $\mathbf{P}$ vanishes in solid solutions of the PbZr$_{1-x}$%
Ti$_{x}$O$_{3}$ (PZT) type.\cite{II99b,II05b} Yet, the notion that the shear
compliance must become large does not provide information on the functional
dependence on temperature, while the exact knowledge of that dependence from
a numerical example is hardly generalizable to other cases.

An intermediate level of analysis has been provided in an attempt to cope
with the transition from T to monoclinic (M) structure occurring at the
morphotropic phase boundary (MPB, namely a nearly vertical boundary in the
composition-temperature phase diagram) of PZT and other PbTiO$_{3}$-based
ferroelectrics.\cite{127,145} These transitions are almost intractable from
the elastic point of view with the required free energy expansion, that must
include at least up to the 8th order of powers of $\mathbf{P}$. Here it is
shown that the same approach can be applied to any case of transition
essentially consisting of a rotation of the direction of the spontaneous
polarization, with little change of its magnitude, and as experimental
verification the classical case of BaTiO$_{3}$ is revisited.

Previous studies of the elastic properties of BaTiO$_{3}$ have been focused
on selected elastic constants, without reference to the rotational
instability at the FE/FE transitions,\cite{BJ58} on the precursor FE
fluctuations and softening in the paraelectric phase,\cite%
{KHI73,CDS78,KKR11,SCN13,Car15} and on the motion of domain walls (DW) and
defects within the FE phases.\cite{CGF94,CWX14}

\section{Experimental}

Two samples from different preparations by conventional mixed-oxide power
techniques were tested. For sample \#1 the starting nominal composition was
BaTiO$_{3}$ $+0.01$ TiO$_{2}$. After mixing (distilled H$_{2}$O), calcining
(1100~$^{\circ }$C, 2h) and fine-milling (distilled H$_{2}$O), the powder
was densified to rectangular bars (about 3~g/cm$^{3}$) and then sintered in
air (1400~$^{\circ }\text{C}$, 1h). The powder X-ray diffraction did not
reveal any trace of impurity phases. The samples had densities of about 92\%
of the theoretical value (6.02 g/cm$^{3}$) and a mean grain size of about 50~%
$\mu \text{m}$. Sample \#1 was cut as a bar of dimensions $33\times
4.2\times 1.1~$mm$^{3}$.

Sample \#2 was prepared from stoichiometric amounts of BaCO$_{3}$ (Aldrich,
99\%) and TiO$_{2}$ (Aldrich, 99.9\%) powders wet-mixed in water using
zirconia media and adding ammonium polyacrylate as dispersant. After
freeze-drying, the mixture was calcined in air for 4~h at 1100~$^{\circ }$C.
The resulting powder was sieved, compacted in bars by means of isostatic
cold pressing and sintered for 2~h at 1450~$^{\circ }$C. The final relative
density of the ceramics was 97\% of the theoretical density. X-ray
diffraction did not reveal any trace of secondary phases within the
detection limit of the technique (1-2 wt\%). The final shape of sample \#2
was $46\times 4.5\times 0.5$~mm$^{3}$, but since not all the sides were cut
after sintering, some irregularities in the shape made the measurement of
the higher frequencies noisy.

The dynamic Young's modulus $E=$ $E^{\prime }+iE^{\prime \prime }$ was
measured by suspending the sample on two thin thermocouple wires in
correspondence with the nodal lines of the first free flexural resonance
mode, almost coinciding with a pair of the nodes of the fifth mode. Silver
paint was applied at the center of one of the faces and the edge, in order
to make electrical contact between the region facing an electrode and the
thermocouple, also acting as ground. The vibration at frequency $f$ is
electrostatically excited by application of an alternate voltage with
frequency $f/2$ to the electrode. The same electrode is part of a resonating
circuit whose high frequency ($\sim 12$~MHz) is modulated by the distance
from the sample, so that the vibration is monitored with a frequency
modulation technique.\cite{135} From the fundamental resonance frequency $%
f_{1}$ of the sample it is possible to deduce the Young's modulus\cite{NB72}
$E\propto f_{1}^{2}$. During a same run also the next odd flexural
vibrations may be excited, whose frequencies ratios with $f_{1}$ are $5.4\ $%
and $13.2$. The data are presented as compliance $s=s^{\prime }-is^{\prime
\prime }=$ $1/E$, normalized to its minimum value $s_{0}$, as $s^{\prime
}\left( T\right) /s_{0}\simeq $ $f_{0}^{2}/f_{1}^{2}\left( T\right) $,
neglecting the changes in cell size with temperature. The elastic energy
loss coefficient $Q^{-1}=s^{\prime \prime }/s^{\prime }$ is measured from
the decay of the free oscillations, if longer than 0.1~s, or from the width
of the resonance peak.

\section{Results}

\begin{figure}[tbh]
\includegraphics[width=8.5 cm]{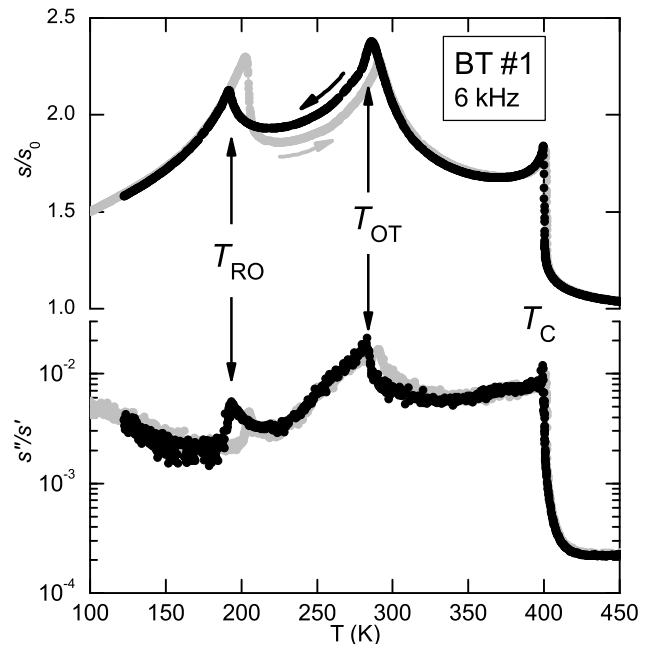}
\caption{Normalized compliance $s^{\prime }$ and elastic energy losses $%
s^{\prime \prime }/s^{\prime }$ of sample BT \#1 measured at 6~kHz during
cooling and heating.}
\label{fig_BT1}
\end{figure}

Figure \ref{fig_BT1} displays the real part of the compliance $s$ and the
losses measured at 6~kHz during heating and cooling of sample BT \#1. All
three phase transitions are clearly visible, similarly to previous
measurements on ceramic samples.\cite{CGF94,CGM03,CWX14,Car15} The
temperatures of the extrema of $s^{\prime }\left( T\right) $ are $T_{\mathrm{%
C}}=399.5$~K during cooling, with a shift of $\Delta T_{\mathrm{C}}=1.5$~K
during heating, $T_{\mathrm{OT}}=286$~K with $\Delta T_{\mathrm{OT}}=5$~K
and $T_{\mathrm{RO}}=191.6$~K with $\Delta T_{\mathrm{RO}}=11$~K. Therefore,
they are all first order, particularly the R/O transition, but outside the
temperature range of the O phase the $s^{\prime }\left( T\right) $ curves
are well reproducible between heating and cooling.

\begin{figure}[tbh]
\includegraphics[width=8.5 cm]{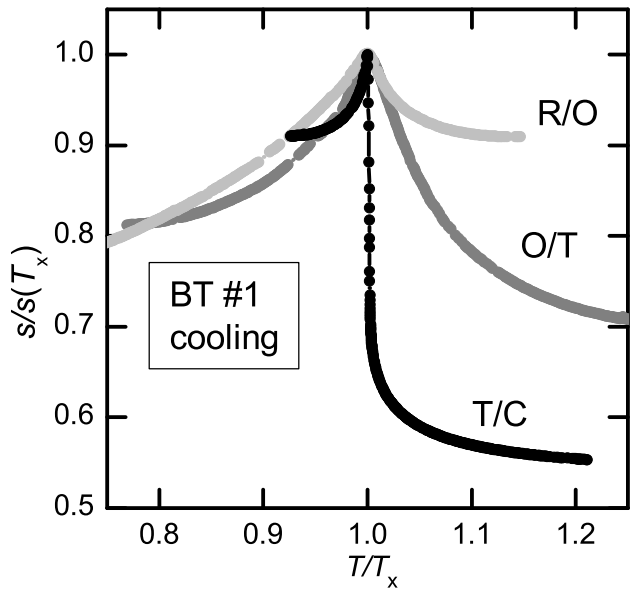}
\caption{Compliance anomalies measured during cooling normalized in
magnitude and temperature.}
\label{fig_norm}
\end{figure}

The most important feature for the present purposes is the fact that $%
s^{\prime }$ has a step at $T_{\mathrm{C}}$ but peaks at $T_{\mathrm{OT}}$
and $T_{\mathrm{RO}}$. The step at $T_{\mathrm{C}}$ is rounded by some
precursor softening in the C-PE phase, that has been amply studied in terms
of fluctuations of the polarization.\cite{KHI73,CDS78,KKR11,SCN13} In
addition, the stiffening just below $T_{\mathrm{C}}$ is mainly related to
domain wall (DW) relaxation, as shown later. On the other hand, the two
anomalies at $T_{\mathrm{OT}}$ and $T_{\mathrm{RO}}$ completely lack the
step and exhibit progressive softening on approaching the transitions from
both high and low temperature. Actually, $s^{\prime }$ makes a jump during
heating through the R/O\ transition, but this is due to large thermal
hysteresis: the R phase remains metastable well into the stability region of
the O phase and the transition appears more abrupt. The different behaviors
at the three transitions are better put in evidence by normalizing the three
anomalies as $s^{\prime }/s^{\prime }\left( T_{x}\right) $ vs $T/T_{x}$ in
Fig. \ref{fig_norm}.

In order to verify a possible dependence of the shapes of the elastic
anomalies on DW relaxation, the measure has been repeated on the longer and
thinner BT \#2, where it was possible to excite also the 3rd and 5th free
flexural modes.

\begin{figure}[tbh]
\includegraphics[width=8.5 cm]{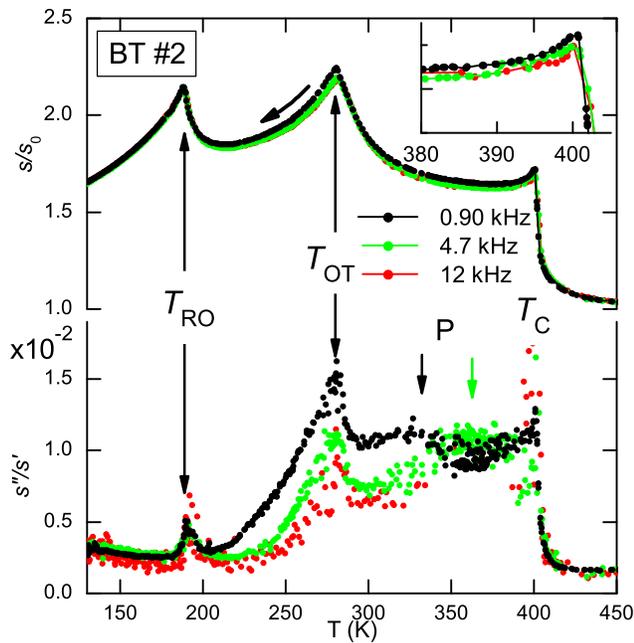}
\caption{Normalized compliance $s^{\prime }$ and elastic energy losses $%
s^{\prime \prime }/s^{\prime }$ of sample BT \#2 measured at 0.9, 4.7 and
12~kHz during cooling.}
\label{fig_BTn2}
\end{figure}

The dependence on frequency is much more evident in the losses than in the
real part, because they are totally due to relaxation of DW, defects and to
fluctuations, while $s^{\prime }$ has a prevalent elastic intrinsic
contribution from the phase transformations. The frequency dispersion of $%
s^{\prime }$ is more evident below $T_{\mathrm{OT}}$ and $T_{\mathrm{C}}$,
in correspondence with the highest values of the losses, and can therefore
be attributed to DW relaxation. The inset shows that at higher frequencies
the upturn of $s^{\prime }$ just below $T_{\mathrm{C}}$ shifts to higher
temperature and also decreases, suggesting that it is totally due to DW
relaxation. A similar feature has been observed more clearly in PZT.\cite%
{127,Cor15} Instead, the rise of $s^{\prime }$ below $T_{\mathrm{OT}}$
cannot be attributed to DW relaxation, because passing from 0.9 to 12~kHz
the amplitude of the losses, and hence of DW relaxation, is halved, while $%
s^{\prime }$ is almost unchanged.

\section{Discussion}

\subsection{Elastic anomaly at a 2nd order phase transition from Landau
theory\label{sect-elastic}}

We remind the main results that can be obtained from the Landau theory of
the phase transitions regarding the elastic anomalies at ferroelectric
transitions, referring to the simplest case of 2nd order transition with
one-dimensional order parameter $P$. In this case the Gibb's free energy is
written as\cite{SL98,Cor15}%
\begin{eqnarray}
G &=&\frac{\alpha }{2}P^{2}+\frac{\beta }{4}P^{4}-\frac{1}{2}%
s_{ij}^{0}\sigma _{i}\sigma _{j}-L_{i}\sigma _{i}P-Q_{i}\sigma _{i}P^{2}~
\label{G} \\
\alpha &=&\alpha ^{\prime }\left( T-T_{\mathrm{C}}\right)  \notag
\end{eqnarray}%
where all the coefficients of the expansion in powers of $P$ and stress $%
\sigma $ are independent of temperature except that of $P^{2}$, which
decreases linearly and becomes negative below $T_{\mathrm{C}}$. The elastic
energy of the symmetric PE phase is expressed in terms of $s_{ij}^{0},$where
$i,j=1-6$ are in Voigt notation ($xx\rightarrow 1,$ $yy\rightarrow 2,$ $%
zz\rightarrow 3,$ $yz\rightarrow 4,$ $xz\rightarrow 5,$ $xy\rightarrow 6$)
and there is summation over repeated indexes. The first coupling term,
linear in both $P$ and $\sigma $ is forbidden by symmetry,
because it must be symmetric under inversion, like the cubic symmetric phase that it also
describes. Since an inversion causes $P\rightarrow -P$ and $\sigma
\rightarrow \sigma $ (stress $\sigma $ and strain $e$ are centrosymmetric
2nd rank tensors), the bilinear term changes sign and is forbidden: it must
be $L=0$. Then the first allowed coupling term is the electrostrictive $%
Q\sigma P^{2}$. The resulting anomalies in the compliance $s_{ij}=\frac{%
de_{i}}{d\sigma _{j}}$, taking into account the variation of the equilibrium
$P$ due to the application of $\sigma $, have been calculated also for more
complicated\cite{II99b,II05b} and general\cite{ST70,SL98} cases. Even though
we have just shown that it must be $L=0$ in the free energy Eq. (\ref{G}) of
a cubic PE phase, in the following discussion it will be useful to see the
effect of the bilinear coupling term on the elastic anomaly, which in the
general case $L\neq 0$ can be written as\cite{SL98,Cor15}
\begin{equation}
s_{ij}\left( T>T_{\mathrm{C}}\right) =s_{ij}^{0}+\frac{L_{i}L_{j}}{\alpha
^{\prime }\left( T-T_{\mathrm{C}}\right) }\;.  \label{s >TC}
\end{equation}%
\begin{gather}
s_{ij}\left( T<T_{\mathrm{C}}\right) =s_{ij}^{0}+  \label{s <TC} \\
+\frac{L_{i}L_{j}}{2\alpha ^{\prime }\left( T_{\mathrm{C}}-T\right) }+\frac{%
\left( L_{i}Q_{j}+L_{j}Q_{i}\right) }{\sqrt{\alpha ^{\prime }\beta \left( T_{%
\mathrm{C}}-T\right) }}+\frac{2Q_{i}Q_{j}}{\beta }~.  \notag
\end{gather}%
Normally, $L=0$ and the only effect of the ferroelectric transition is a
steplike softening below $T_{\mathrm{C}}$ of magnitude $2Q^{2}/\beta $, as
observed at $T_{\mathrm{C}}$ in BaTiO$_{3}$ and Ti-rich PZT.\cite{127}
Instead, the bilinear coupling alone causes a divergence exactly of the
Curie-Weiss type both above and below $T_{\mathrm{C}}$, of magnitude $\propto L^{2}/\alpha
^{\prime }$. This type of behavior is indeed observed when the PE phase is
piezoelectric; for example, a much studied case is the $c_{66}$ elastic
constant of KDP.\cite{LG70} Notice also that, already in the simple case of
a one-dimensional order parameter, the two types of anomalies are not simply
additive below $T_{\mathrm{C}}$, since there is a mixed term, but above $T_{%
\mathrm{C}}$ only the Curie-Weiss term from bilinear coupling is present.

\subsection{Relationship between dielectric, elastic and piezoelectric
responses}

Before discussing further the nature of the softenings at the FE
transitions, we would like to stress that they participate to the
enhancement of the piezoelectric effect. A particularly simple relationship
in this respect can be obtained starting from the original model\cite{MM48}
of ferroelectricity in terms of thermodynamic equilibrium of the possible
orientations of the spontaneous polarization in each cell, producing a high
frequency dielectric relaxation with amplitude $\Delta \chi $, and also
considering the anelastic relaxation with amplitude $\Delta s$ (enhancement
of the compliance) from the associated anisotropic strain (\textit{e.g.}
tetragonality in a T-FE phase).\cite{Cor15} Then, the amplitude of the
piezoelectric effect can be expressed as\cite{Cor15} (see Refs. %
\onlinecite{NH65,Dam06} for an analogous expression for the relaxation of
non interacting point defects)
\begin{equation}
d=\sqrt{\Delta \chi \Delta s}\simeq \sqrt{\varepsilon \Delta s}
\end{equation}%
where the magnitude of the dielectric constant can be almost completely
associated with the FE phase, while the compliance $s^{0}$ in the
paraelectric phase in general cannot be neglected with respect to the
enhancement $\Delta s$ in the FE phase. The above formula expresses the fact
that the piezoelectric effect is made on an equal basis of a change in
polarization and in strain, the first producing an enhancement of the
dielectric susceptibility and the latter of the elastic compliance.

\subsection{Nearly linear coupling between polarization rotation and shear
strain}

At the middle of the MPB of PZT a large softening reminiscent of the
Curie-Weiss term is observed, that has been associated with the T/M
transition, with the M phase intermediate between T and R.\cite{127,145} In
order to explain the peaked softening, anomalously large and extended in
temperature, it has been observed that the main effect of the T/M transition
is a rotation of $\mathbf{P}$ away from $\left( 001\right) ,$ and the shear
strains $e_{4}$ and $e_{5}$ are almost linearly coupled to the rotation
angle, so that the elastic compliances $s_{44}$ and $s_{55}$ are expected to
diverge as Eqs. (\ref{s >TC},\ref{s <TC}) with $L\neq 0$.\cite{127,145}

We reproduce here the argument for the case of the T/O transition of BaTiO$%
_{3}$ and observe that it is more general than the case of the T/M
transition with little change of the orientation of $\mathbf{P}$, in fact it
applies to any transition where the main result is a rotation of $\mathbf{P}$%
.

\begin{figure}[tbh]
\includegraphics[width=7.5 cm]{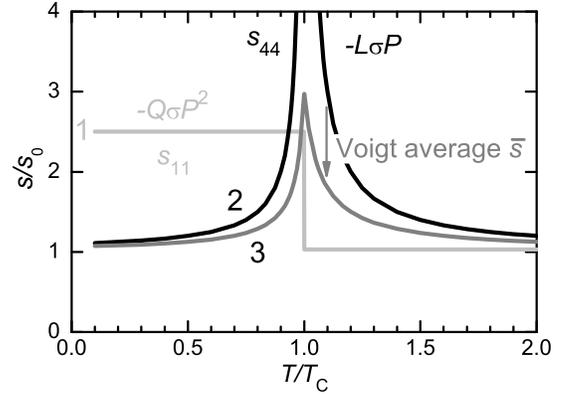}
\caption{Polarization at the T/O transition; $e_{xz}$ is the strain coupled
to the polarization rotation. The magnitudes of $\mathbf{P}$ are according
to the eight site model of BaTiO$_{3}$, and also shown are the eight
off-center positions of Ti.}
\label{fig_pol-rot}
\end{figure}

\begin{figure}[tbh]
\includegraphics[width=7.5 cm]{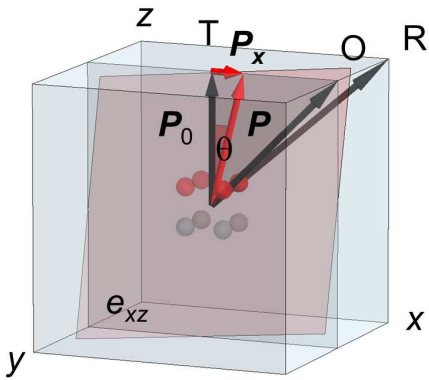}
\caption{Expected anomalies in the compliance at a phase transition from the
linear and the quadratic coupling of stress to the order parameter (Eq. (%
\protect\ref{G})). (1) step-like softening from the linear-quadratic
coupling; (2) Curie-Weiss-type softening from bilinear coupling, (3)
simulation of polycrystalline average of curve (2) with other constant
compliances.}
\label{fig_s-vs-T}
\end{figure}

Looking at Fig. \ref{fig_pol-rot}, we express $\mathbf{P}$ in terms of $%
\mathbf{P}_{0}$ in the T phase and of the transverse components $P_{x}=$ $%
P_{0}\sin \theta _{x}$, $P_{y}=$ $P_{0}\sin \theta _{y}$, where $\theta _{x}$
and $\theta _{y}$ are the rotation angles of $\mathbf{P}$ away from the T
direction in the $xz$ and $yz$ planes. Then, the electrostrictive coupling
term permitted by symmetry, $G_{c}=-Q_{ijk}\sigma _{i}P_{j}P_{k}$, becomes
\begin{eqnarray}
-G_{c} &=&Q_{11}\left[ \sigma _{1}P_{x}^{2}+\sigma _{2}P_{y}^{2}+\sigma
_{3}P_{0}^{2}\right] +  \notag \\
&&+Q_{12}\left[ \left( \sigma _{1}+\sigma _{2}\right) P_{0}^{2}+\sigma
_{3}\left( P_{x}^{2}+P_{y}^{2}\right) \right] + \\
&&+Q_{44}\left[ \left( \sigma _{4}P_{x}+\sigma _{5}P_{y}\right) P_{0}+\sigma
_{6}P_{x}P_{y}\right]  \notag
\end{eqnarray}%
where the Voigt notation is used also for the last pair of indexes of $%
Q_{ijk}$. Until we disregard the variation of $P_{0}\ $compared to that of
the transverse components $P_{x},P_{y}$, the latter behave as the order
parameters of the T/O transition. In addition, as long as $P_{x},P_{y}$ $\ll
P_{0}$, we also neglect the terms quadratic in $P_{x},P_{y}$, consider as
constants, or producing steplike elastic anomalies those quadratic in $P_{0}$%
, and focus on
\begin{equation}
G_{c}\simeq -Q_{44}\left( \sigma _{4}P_{x}+\sigma _{5}P_{y}\right) P_{0}~.
\label{Gc-bilin}
\end{equation}%
that is linear both in the shear stresses $\sigma _{4}$ and $\sigma _{5}$
and in the order parameters $P_{x}$ and $P_{y}$. We are exactly in the
situation of Eqs. (\ref{G}-\ref{s >TC}) with $L=-Q_{44}P_{0}$ yielding
\begin{equation}
s_{44}=s_{55}=s_{44}^{0}+\frac{Q_{44}^{2}P_{0}^{2}}{\alpha ^{\prime }}%
\left\{
\begin{array}{c}
\frac{1}{T-T_{0}}\;\left( T>T_{0}\right) \\
\frac{1}{2}\frac{1}{T_{0}-T}\;\left( T<T_{0}\right)%
\end{array}%
\right. ~,  \label{s-CW}
\end{equation}%
where $T_{0}=T_{\mathrm{OT}}$ for the O/T transition. Apart from the
approximations done, including that of a 2nd order transition, the
divergence in ceramic samples is trimmed down by the fact that one measures
a combination of elastic constants, of which only $c_{44}=1/s_{44}$ and $%
c_{55}=1/s_{55}$ vanish at $T_{0}$, while the others keep the effective
modulus finite. The appearance of this type of elastic anomaly is
demonstrated in Fig. \ref{fig_s-vs-T} for the two main types of elastic
responses. Curve (1) is Eq. (\ref{s <TC}) with $2Q^{2}/\beta =1.5$, $s^{0}=1$%
, corresponding for example to $s_{33}$ at $T_{\mathrm{C}}$; curve (2) is
Eq. (\ref{s-CW}) with $Q_{44}^{2}P_{0}^{2}/\alpha ^{\prime }=0.1$, $s^{0}=1$%
; curve (3) is a polycrystalline average of the latter response with other
constant compliances, the Voigt type average of $2/3$ of curve (2) with
another constant compliance $s_{1}=1/3$: $\overline{s}=\left(
s_{1}^{-1}+s_{44}^{-1}\right) ^{-1}$.

Notice that the condition for nearly isotropic free energy, that is not
applicable in the simple free energy Eq. (\ref{G}) but has been amply
discussed,\cite{II99b,II05b,Kha10} makes it possible the rotational
instability of the polarization, and hence the T/O transition, but the above
expression of the diverging compliance contains additional information
related to the piezoelectric response. In fact, the magnitude of the effect
is proportional to: \textit{i)} the square of the electrostrictive coupling
involved in the transition; \textit{ii)} the square of the magnitude of the
polarization, and \textit{iii)} the reciprocal curvature or flatness of the
free energy minimum near the transition. This means that the magnitude of
the maximum of the compliance is closely correlated to that of the
piezoelectric effect, and, considering that the elastic measurements on
unpoled ceramics at kHz and lower frequencies are insensitive to free
charges, make this type of measurement a valid tool for investigating solid
solutions where large piezoelectric responses from the rotation of the
polarization are expected.

The assumption leading to Eq. (\ref{s-CW}) of a transition with continuous
rotations of the polarization and no change of its magnitude is akin to the
introduction of a reduced free energy depending only on the polarization
direction in order to study the equilibrium ranges of the possible FE
phases, including the monoclinic ones.\cite{VC01} In another approach to the
same problem, Hudak\cite{Hud08} redefined the order parameter as $\mathbf{P=}
$ $P\sin \theta $, as above. The present approach is to neglect the
variation of $P$ and focus on that of $\sin \theta \sim \theta $, since the
analysis of the possible phases has already been done in various
approximations,\cite{HFM89,VC01,Hud08} and we are only interested in the
effect of the polarization rotation, hence of $\theta $, on the elastic
response. The presence of additional components of the order parameter will
introduce additional features to the elastic anomaly, like a jump.

\subsection{The case of BaTiO$_{3}$}

In the Results it has been shown that the peaks in the reciprocal Young's
modulus of ceramic BaTiO$_{3}$ are intrinsic and not a result of DW
relaxation. To the same conclusion lead the measurements made with the
torsional pendulum over a broad range of lower frequencies\cite{CGM03} ($%
0.01-1$~Hz): the peaked softenings are independent of frequency, even though
the losses contain large relaxational contributions. We do not attempt fits
of the $s^{\prime }\left( T\right) $ curves in terms of Eq. (\ref{s-CW})
because they would be of limited significance, considering the
polycrystalline average, the overlapping of the effects of two rotational
transitions of the polarization, the approximation of 2nd order transition
leading to Eq. (\ref{s-CW}), and its limited validity below the rotational
transition, when $P_{x}$ becomes comparable to $P_{0}$. Nonetheless, it is
clear that the peaked softenings at $T_{\mathrm{OT}}$ and also $T_{\mathrm{RO%
}}$ are of the type of Eq. (\ref{s-CW}).

In BaTiO$_{3}$ the schematization of the spontaneous polarization changing
direction but not magnitude at the FE/FE transition is not strictly valid.
In fact, the transitions can also be schematized in terms of progressive
localization of the Ti ions in the $\left\langle 111\right\rangle $
off-center positions belonging to a same cube face (T), edge (O) and a
single off-center site (R). The experimental evidence supporting such an
order-disorder character comes for example from EXAFS experiments \cite%
{RSV98} and from the observation of non null electric field gradient at the
Ti site also in the cubic phase.\cite{Bli04} In the model of progressive
localization of Ti over the eight off-center sites, the magnitude of the
polarization should increase of a factor of $\sqrt{2}$ in the O phase with
respect to the T phase, and of $\sqrt{3}$ in the R phase. In practice, the
change in the magnitude of $\mathbf{P}$ during the various transitions is
smaller and nearly continuous,\cite{KLB93} and appears as a consequence of
the FE/PE transition; only the $P_{z}$ projection on the $c$ axis has jumps
at the first order  FE/FE transitions, validating the approximation of their
order parameter as the rotation angle or transverse component of $\mathbf{P}$%
. In addition, the transitions in BaTiO$_{3}$ are generally treated as
displacive\cite{Bli04} or at least the displacive and order-disorder
characters are seen to coexist.\cite{BBV09b} Therefore, the assumption that
the FE/FE transitions in BaTiO$_{3}$ consist mainly of rotations of the
polarization is correct. The assumption of the smallness of the transverse
components of $\mathbf{P}$ producing its rotation is certainly valid and
explains the rise of the compliance when approaching $T_{\mathrm{OT}}$ and $%
T_{\mathrm{RO}}$ from above, where these components only fluctuate with null
mean value. The strong first order nature of the FE-FE transitions may
invalidate the approximation of small magnitude of the order parameter below
their temperatures $T_{\mathrm{OT}}$ and $T_{\mathrm{RO}}$, but this does
not exclude the enhancement of the elastic compliance of the Curie-Weiss
type, as experiments show; simply, in these temperature regions one cannot
exploit the approximation of nearly bilinear stress-polarization coupling
and has to carry out the full calculations.

\subsection{Generality of the shear softening at the rotational transition
of the polarization}

As also shown in a recent review,\cite{Cor15} PZT at the MPB is
an outstanding case of maximum of the compliance associated with a
rotational instability of the polarization, but there are other
perovskite solid solutions that conform to the present analysis:\
that of PbTiO$_{3}$ with the relaxor ferroelectric
PbMg$_{1/3}$Nb$_{2/3}$O$_{3}$,\cite{AJP05} the
pseudobinary $(1-x)$Ba(Ti$_{0.8}$Zr$_{0.2}$)O$_{3}-x$(Ba$_{0.7}$Ca$_{0.3}$%
)TiO$_{3}$ with $x\sim $ 0.5,\cite{XZB11,DBB12,CCD14b} that can be seen as a
tuning close to room temperature of the sequence of transitions of BaTiO$%
_{3} $, of K$_{1-x}$Na$_{x}$NbO$_{3}$ (KNN) and some KNN-based compounds\cite%
{GRZ15} and possibly (Na$_{1/2}$Bi$_{1/2}$)$_{1-x}$Ba$_{x}$TiO$_{3}$
(NBT-BT).\cite{139} The case of NBT-BT is uncertain; indeed a broad peak in
the elastic compliance develops on approaching the MPB composition $x\sim
0.06$ separating R and T phases, but there is scarcity of elastic data at
higher Ba content, and it seems that the softening is connected to a
strain-glass transition, involving short range ordering of strain rather
than polarization rotation.\cite{YSJ13}

Instead, the K$_{1-x}$Na$_{x}$NbO$_{3}$ system with $x\simeq 0.5$ is
particularly interesting in the present context. It exhibits a C-PE/T-FE
transition at $T_{\mathrm{C}}=670$~K, ~followed by a transition to O or M
phase with rotation of $\mathbf{P}$ at $T_{\mathrm{OT}}=470$~K, and below $%
T_{\mathrm{OT}}$ there is a vertical MPB in the $x-T$ phase diagram between
O and M phases differing in octahedral tilting rather than in the direction of $%
\mathbf{P}$.\cite{BTZ09} The piezoelectric coupling $d$ presents a sharp
maximum right at $T_{\mathrm{OT}}$, the temperature of the rotational
instability, and at the MPB is lower then within the T phase,\cite{GLV11}
confirming the essential role of the rotational instability of $\mathbf{P}$
compared to the presence of lattice disorder and MPBs.\cite{MMP13} The
latter enhance $d$ in as much as they favor the rotational instability and
extend its temperature range. Accordingly, Resonant Ultrasound Spectroscopy
measurements reveal that the bulk modulus of KNN undergoes minor changes at $%
T_{\mathrm{OT}}$, while the shear modulus behaves similarly to the Young's
modulus of PZT close to the MPB and of the other systems just mentioned:
step at $T_{\mathrm{C}}$ and minimum of the type of Eq. (\ref{s-CW}) at $T_{%
\mathrm{OT}}$.\cite{YLC11} Evidently, while the bulk modulus has no
contributions from the volume-conserving shears, the shear modulus has an
important contribution from the step of $c_{11}-c_{12}$ at $T_{\mathrm{C}}$
and from the Curie-Weiss-type softening of $c_{44}$ and $c_{55}$ at $T_{%
\mathrm{OT}}$. The latter must be responsible for the peak in the
piezoelectric coupling.\cite{GLV11}

We emphasize that the approximation of quasi-bilinear coupling between shear
and rotation of the polarization does not depend on the fact that the rotational
instability is obtained by changing only temperature, as in  undoped BaTiO$_3$,
or also the composition, as in the PbTiO$_3$-based solid solutions.
The approximation holds until the order parameter is close to a rotation of the
polarization and its mean value is small. The condition of smallness of the order
parameter is certainly satisfied above the temperature of the transition, where
its mean value is null, and possibly also in a temperature range below the transition
temperature, unless a strong jump in the direction of the polarization occurs due
to a marked first order character of the transition. In the latter cases one cannot
avoid working out the full calculations.

\section{Conclusion}

The relationship between enhancement of the piezoelectric coupling and
rotational instability of the polarization with consequent enhancement of
the shear elastic compliance has been discussed. The previous analysis of
the tetragonal/monoclininc transition at the morphotropic phase boundary of
PZT\cite{127,145} has been extended to the general case of transitions with
rotation of the electric polarization, and as experimental verification the
elastic response of BaTiO$_{3}$ has been revisited. The important point is
that the rotation of the polarization is almost linearly coupled to the
corresponding shear strain, leading to an enhancement of the shear
compliance of the type $1/\left\vert T-T_{0}\right\vert $, whereas normally
ferroelectric transitions cause steps in the elastic constants. Particularly
the enhancement at $T>T_{0}$, is obtained within the usual Landau theory
without introducing the fluctuations. Therefore, this divergence is an
intrinsic robust feature to be found whenever there is a rotational
instability of the polarization, and it substantially contributes to the
enhancement of the piezoelectric coefficient.

It follows that the measurement of the elastic properties is very
informative in the study of the phase diagrams of solid solutions designed
for obtaining giant piezoelectricity, particularly in possible cases where
extrinsic defects and the associated free charges obscure the ferroelectric
and piezoelectric responses.

\begin{acknowledgments}
F.C. thanks P.M. Latino (CNR-ISC) for his technical assistance and
improvements in the electronics for detecting the sample vibration.
\end{acknowledgments}

\bibliographystyle{apsrev}
\bibliography{/articles/refs}

\end{document}